\documentclass[aps,showpacs,nofootinbib]{revtex4}
\usepackage{amsmath} 
\usepackage{amssymb}
\usepackage{graphicx} 
\usepackage{color} 
\usepackage{mathtext}
\usepackage{epstopdf}

\begin{document}
\newcommand{\blue}[1]{\textcolor{blue}{#1}}
\newcommand{\new}{\blue}
\newcommand{\green}[1]{\textcolor{green}{#1}}
\newcommand{\modif}{\green}
\newcommand{\red}[1]{\textcolor{red}{#1}}
\newcommand{\attention}{\red}


\title{Energy dependence of cross sections in proton--proton and antiproton--proton collisions}

\author{V. A. Okorokov} \email{VAOkorokov@mephi.ru; Okorokov@bnl.gov}
\affiliation{National Research Nuclear University MEPhI (Moscow
Engineering Physics Institute), Kashirskoe highway 31, 115409
Moscow, Russia}

\date{\today}

\begin{abstract}
\noindent {\bf Abstract}---Energy dependence of global scattering
parameters, mostly of total cross section, is studied for
proton--proton and antiproton--proton collisions. Results are
presented for physical analysis updated with taken into account
the recent data from accelerator experiments as well as from
cosmic ray measurements. The analytic parameterizations suggested
within Axiomatic Quantum Field Theory (AQFT) provide the
quantitative description of energy dependence of global scattering
parameters for rather wide energy range. Detailed scan on low
boundary of the fitting range for energy dependence of global
scattering parameters allows the observation of the onsets for
regions in which Pomeranchuk theorem and / or Froissart--Martin
one is valid. It is obtained that global scattering parameters
show the behavior corresponded to any formulations of Pomeranchuk
theorem and closed to (modified) Froissart--Martin limit in
functional sense in multi-TeV energy region. Bosonic condensation
is considered as one of the possible dynamical mechanisms which
would be provide the total cross section approaches to (modified)
Froissart--Martin limit at quantitative level but not functionally
only.
\end{abstract}


\maketitle

\section{Introduction}\label{sec:1}

The large-scale projects of hadronic colliders of next generation
assume the measurements for proton--proton ($pp$) interactions at
collision energies up to $\sqrt{s} \sim 100$ TeV in order to
magnitude \cite{arXiv-1507.03224-2015,EPJC-79-474-2019}.
Phenomenological study of cross sections as one the most important
global quantities for (anti)proton--proton ($\bar{p}p$, $pp$)
scattering at very high $\sqrt{s}$ can be helpful for search for
the onset of the asymptotic regime in energy dependence of cross
sections and related parameters as well as for the establish of
more precision and reliable criteria for asymptotic domain in both
mathematical and physical senses. The rigorous theorems and
relations between global scattering parameters are crucially
important for high energy physics, particularly for the asymptotic
energy domain. In this paper, two of the most outstanding among
them, namely the Pomeranchuk theorem \cite{JEPT-7-499-1958} and
the Froissart or, perhaps, more correctly Froissart--Martin one
\cite{PR-123-1053-1961,PR-129-1432-1963} are considered, both
concerning $pp$ and $\bar{p}p$ scattering processes. Importance of
the asymptotic theorems for high energy physics is in the ability
for direct and model-independent verification of a most general
and basic principles which are base for Quantum Field Theory (QFT)
at all. However, the theory, while predicting the behavior of some
global scattering parameters for particle--particle and
antiparticle--particle scattering processes, provides no
indications and / or estimates, even by order of magnitude, of the
energies at which the asymptotic region begins. As it is known a
limit in the mathematical sense is only indicated. Various
observables and its combinations are used for the study of
asymptotic regime
\cite{Okorokov-IJMPA-33-1850077-2018,Okorokov-PAN-81-508-2018,Okorokov-PAN-82-134-2019,
Okorokov-CPC-46-083105-2022}. Estimations for low boundary on
$\sqrt{s}$ for onset of asymptotic regime depend on model and / or
choice of a global observable used for consideration. Such
estimations can differ (very) significantly from each other.
Furthermore the definition of term ''asymptote" itself in physical
sense for global scattering parameters or ''truly asymptotic
region" is non-trivial \cite{Okorokov-IJMPA-33-1850077-2018}. As
consequence, the study of global scattering parameters at high and
very high collision energies, in particular, for direct rigorous
verification of asymptotic theorem is rather complex problem in
various sub-fields (theory, phenomenology, data analysis,
experiment) of modern physics of strong interaction. Therefore,
any estimates, even qualitative ones in order of magnitude, of the
lower energy bound for the onset of the asymptotic region seem
important. Such information can be obtained using phenomenological
models and physical analysis of experimental data. Perhaps, such
studies may be actual both for a better understanding of the
dynamics of strong interactions at large distances (small-angle
scattering) in wide energy domain and for the development of
physics research programs for future accelerator facilities. In
the paper the Axiomatic Quantum Field Theory (AQFT) approach is
used for phenomenological study of energy dependence of global
scattering parameters, mostly, of total cross sections.

\section{Asymptotic relations and AQFT model}\label{sec:2}

The Pomeranchuk theorem allows the two formulations
\cite{Okorokov-CPC-46-083105-2022}. The original version assumes
that if the forward elastic scattering amplitude
$\mathcal{F}^{\pm}(s)$, where the plus sign is for $pp$ and the
minus one for $\bar{p}p$ interaction\footnote{As indicated above,
the theorem is valid for general case particle--particle and
antiparticle--particle processes but here the formulation is
adopted for specific reactions under study.} for brevity, have no
oscillations at $s \to \infty$ and they are mostly imaginary
$\displaystyle \lim_{\varepsilon \to \infty} [\mathrm{Re}
\mathcal{F}^{\pm}] / [\mathrm{Im} \mathcal{F}^{\pm}\ln
\varepsilon]=0$ with $\varepsilon \equiv s/s_{0}$, $s_{0}=1$
GeV$^{2}$, then the difference between $pp$ and $\bar{p}p$ total
cross sections tend to zero in asymptotic region
\begin{eqnarray}\label{eq:2.1}
\Delta_{\footnotesize\mbox{t}}(s)\equiv
|\sigma_{\scriptsize{\mbox{t}}}^{+}-\sigma_{\scriptsize{\mbox{t}}}^{-}|
\rightarrow 0, ~\mathrm{at}~ s \to \infty.
\end{eqnarray}

\noindent This relation was derived under the assumption of
asymptotically constant cross sections
$\sigma_{\scriptsize{\mbox{t}}}^{\pm} \to
C^{\pm}_{\scriptsize{\mbox{P}}}$ at $s \to \infty$, where
$C^{\pm}_{\scriptsize{\mbox{P}}} \ne 0$ are some constants and
$|C^{+}_{\scriptsize{\mbox{P}}}-C^{-}_{\scriptsize{\mbox{P}}}| \to
0$ at $s \to \infty$ as the condition for correctness of
(\ref{eq:2.1}).

The modified, or generalized for increasing cross sections,
formulation of the theorem means that
\begin{eqnarray}\label{eq:2.2}
R_{\footnotesize\mbox{t}}^{-/+}(s) \equiv
\sigma_{\footnotesize\mbox{t}}^{-}(s) /
\sigma_{\footnotesize\mbox{t}}^{+}(s) \rightarrow 1, ~\mathrm{at}~
s \rightarrow \infty.
\end{eqnarray}

\noindent This asymptote was proved if
$\sigma_{\scriptsize{\mbox{t}}}^{\pm}$ grow with energy and this
behaviour of $\sigma_{\scriptsize{\mbox{t}}}^{\pm}(s)$ is just
observed in experimentally attainable high-energy domain.

As it is known the Froissart--Martin theorem is formulated as
inequality (upper bound) and it states that any total cross
section cannot grow faster than $\ln^{2}\varepsilon$. There are
two various views of inequality within the Froissart--Martin
theorem in different publications\footnote{Future discussion will
demonstrate that this discrepancy in the formulation of
Froissart--Martin theorem will not affect on the physical
conclusions based on the available experimental databases.}. It is
possible the conditional inequality \cite{Bogolubov-book-1990-1}
and absolute one \cite{PRD-80-065013-2009}; the corresponding
mathematical views of Froissart--Martin theorem are
\begin{subequations}\label{2.3}
\begin{equation}\label{2.3.1}
\sigma_{\scriptsize{\mbox{t}}} \leq
\sigma_{\scriptsize{\mbox{FM}}}, ~\mathrm{at}~ s \to \infty;
\end{equation}
\begin{equation}\label{2.3.2}
\sigma_{\scriptsize{\mbox{t}}} < \sigma_{\scriptsize{\mbox{FM}}},
~\mathrm{at}~ s \to \infty;
\end{equation}
\end{subequations}

\noindent According to the study \cite{PRD-80-065013-2009} the new
bound from Froissart--Martin theorem or, equivalently, modified
Froissart--Martin limit is
$\sigma_{\scriptsize{\mbox{FM}}}=C^{(m)}_{\scriptsize{\mbox{FM}}}\ln^{2}\varepsilon$,
$C^{(m)}_{\scriptsize{\mbox{FM}}} \equiv \pi / 2m^{2}_{\pi}
\approx 31.4$ mb and $m_{\pi}$ is the pion mass
\cite{PRD-110-030001-2024}. The modified limit is twice smaller
than that from original studies \cite{NCA-42-930-1966} based on
the principles of AQFT \cite{Okorokov-CPC-46-083105-2022} due to
relation between constant terms
$C_{\scriptsize{\mbox{FM}}}=2\,C^{(m)}_{\scriptsize{\mbox{FM}}}$
for original and modified limits. The (modified) limit does not
depend on the type of collisions.

The theorems of Pomeranchuk and Froissart--Martin are independent
from each other mathematically and physically. In general, the
energy domains can be different for the validity of each of these
asymptotic theorems. Therefore, the more correct and careful
statement is the reach of the energy domain for validity of
certain theorem or asymptotic region / regime for fixed observable
with point of view of certain theorem if any. But similar
observation will remain the problem of reach of ''truly asymptotic
region" still open and without rigorous quantitative answer.

The analytic parameterizations were proposed within AQFT
\cite{Okorokov-IJMPA-25-5333-2010} in order to describe the energy
dependence of the most known global scattering parameters, namely,
$\sigma_{\scriptsize{\mbox{t}}}^{\pm}$ and $\rho$-parameter
defined as the ratio $\rho^{\pm}=\mathrm{Re} \mathcal{F}^{\pm} /
\mathrm{Im} \mathcal{F}^{\pm}$ in $pp$ and $\bar{p}p$ scattering.
The following form of the parameterizations
\begin{eqnarray}
\label{eq:2.4}
\sigma_{\scriptsize{\mbox{t}}}^{\pm}(s)&=&\frac{\delta_{\pm}}
{\varepsilon^{6+\gamma_{\pm}}}+\frac{\beta_{\pm}}{\varepsilon^{\alpha_{\pm}}}
\,\sigma_{\scriptsize{\mbox{FM}}}, \\
\label{eq:2.5}
\rho^{\pm}(s)&=&\frac{1}{\sigma_{\scriptsize{\mbox{t}}}^{\pm}}\biggl\{\frac{k}{\varepsilon}
+\frac{\pi}{2}\biggl[-\frac{\delta_{\pm}(6+\gamma_{\pm})}
{\varepsilon^{6+\gamma_{\pm}}}+\frac{\beta_{\pm}}{\varepsilon^{\alpha_{\pm}}}\,
\sigma_{\scriptsize{\mbox{FM}}} \biggl(\frac{2}{\ln
\varepsilon}-\alpha_{\pm}\biggr)\biggr] \pm
\frac{\pi}{4}[\sigma_{\footnotesize\mbox{t}}
^{+}-\sigma_{\footnotesize\mbox{t}}^{-}]\biggr\}
\end{eqnarray}

\noindent is more useful for the present study. Here the factor
$\sigma_{\scriptsize{\mbox{FM}}}$ corresponding to the modified
Froissart--Martin limit \cite{PRD-80-065013-2009} is singled out
in clear view.

The relations between free parameters are shown in Table
\ref{tab:1} at which various formulations of Pomeranchuk theorem
are valid within AQFT approach. In most part of cases formulations
(\ref{eq:2.1}), (\ref{eq:2.2}) do not allow numerical estimations
of parameters $\alpha$, $\beta$, $\gamma$, $\delta$ at which
Pomeranchuk theorem is valid.

\begin{table}[t!]
\centering \caption{Relations for AQFT parameters for validity of
Pomeranchuk theorem.}\label{tab:1}
\begin{tabular}{cccc} \hline
\multicolumn{1}{c}{$\alpha$} & \multicolumn{1}{c}{$\beta$} &
\multicolumn{1}{c}{$\gamma$} & \multicolumn{1}{c}{$\delta$}
\\\hline
\multicolumn{4}{c}{Original formulation} \\
\hline $\alpha_{+}=\alpha_{-} \equiv
\alpha_{\scriptsize{\mbox{P}}}$ & $\beta_{+}=\beta_{-} \equiv
\beta_{\scriptsize{\mbox{P}}}$ & $\gamma_{+}=\gamma_{-} \equiv
\gamma_{\scriptsize{\mbox{P}}} > -6$ & $\delta_{\pm}$ are finite \\
\cline{3-4}\rule{0pt}{10pt} & & $\gamma_{+}=\gamma_{-} \equiv
\gamma_{\scriptsize{\mbox{P}}} = -6$ & $\delta_{+}=\delta_{-}
\equiv \delta_{\scriptsize{\mbox{P}}}$ \\
\hline
\multicolumn{4}{c}{Generalized formulation} \\
\hline $\alpha_{+}=\alpha_{-} \equiv
\alpha_{\scriptsize{\mbox{P}}}$ & $\beta_{+}=\beta_{-} \equiv
\beta_{\scriptsize{\mbox{P}}}$ & $\gamma_{+}=\gamma_{-} \equiv
\gamma_{\scriptsize{\mbox{P}}} > -6$, & $\delta_{\pm}$ are finite \\
 & &
$\gamma_{\scriptsize{\mbox{P}}}+6 \geq \alpha_{\scriptsize{\mbox{P}}}$ & \\
\cline{2-4} & $\beta_{\pm}$ are finite & $\gamma_{+}=\gamma_{-}
\equiv \gamma_{\scriptsize{\mbox{P}}} = -6$ &
$\delta_{+}=\delta_{-}
\equiv \delta_{\scriptsize{\mbox{P}}}$ \\
\hline
\end{tabular}
\end{table}

The observables $\sigma_{\scriptsize{\mbox{t}}}^{\pm}$ are
characterized by the one energy dependence (\ref{eq:2.4}) in
functional sense. Thus the (modified) Froissart--Martin limit is
reached at the equal values of AQFT parameters for $pp$ and
$\bar{p}p$ collisions. These asymptotic values with point of view
of the Froissart--Martin theorem are shown in Table \ref{tab:2}.

\begin{table}[t!]
\centering \caption{Values of AQFT parameters for modified
Froissart--Martin limit.}\label{tab:2}
\begin{tabular}{cccc} \hline
\multicolumn{1}{c}{$\alpha_{\scriptsize{\mbox{FM}}}$} &
\multicolumn{1}{c}{$\beta_{\scriptsize{\mbox{FM}}}$} &
\multicolumn{1}{c}{$-\gamma_{\scriptsize{\mbox{FM}}}$} &
\multicolumn{1}{c}{$\delta_{\scriptsize{\mbox{FM}}}$}
\\\hline\rule{0pt}{12pt}
0.0 & 1.0 & 6.0 & 0.0\\
\hline
\end{tabular}
\end{table}

It is seen from Tables \ref{tab:1}, \ref{tab:2} the reach of the
(modified) Froissart--Martin limit by
$\sigma_{\scriptsize{\mbox{t}}}^{\pm}$ leads to the validity of
Pomeranchuk theorem in any formulations within AQFT. But, in
general, the opposite statement is not correct.

The asymptotic theorems considered here establish some relations
for total cross sections but they do not provide any additional
information and / or asymptotic relations for $\rho$-parameter.
From the request of mostly imaginary $\mathcal{F}^{\pm}(s)$ at $s
\to \infty$ it only consequents $\displaystyle
\rho^{\pm}=\mathrm{Re} \mathcal{F}^{\pm} / \mathrm{Im}
\mathcal{F}^{\pm} \ll 1$ in the energy domain of validity of
Pomeranchuk theorem. Within AQFT approach the $\rho$-parameter as
well as total cross section is equal for $pp$ and $\bar{p}p$
collisions in the energy domain of (modified) Froissart--Martin
limit and $\rho$ reaches the asymptotic value
$\rho_{\scriptsize{\mbox{FM}}}=\pi \ln^{-1}\varepsilon$ because
$\kappa/\varepsilon \to 0$ at $\varepsilon \to \infty$ at any
finite $\kappa$. That asymptotic level for $\rho$ agrees with
results for $\left.\rho(s)\right|_{s \to \infty}$ obtained with
help of the crossing property, 1st order derivative dispersion
relations and optical theorem \cite{Okorokov-CPC-46-083105-2022}.

\section{Experimental data and fit procedure}\label{sec:3}

The main global physical quantities characterizing the elastic
scattering process of two particles are the well-known total cross
section ($\sigma_{\footnotesize\mbox{t}}$) and the ratio of the
real to the imaginary part of the forward scattering amplitude
($\rho$ parameter). Therefore the set of global parameters
$\mathcal{G}_{1}=\{\mathcal{G}_{1}^{i}\}_{i=1}^{4}=
\{\sigma_{\scriptsize{\mbox{t}}}^{+},\sigma_{\scriptsize{\mbox{t}}}^{-},
\rho^{+},\rho^{-}\}$ is under study for elastic $pp$ and
$\bar{p}p$ scattering below. The set $\mathcal{G}_{1}$ contains
only observables that are independent of each other as well as
directly measured in experiments.

The choice of the ''standard" database for the present work was
described in details elsewhere
\cite{Okorokov-IJMPA-25-5333-2010,Okorokov-IJMPA-32-1750175-2017}.
The experimental database for $\mathcal{G}_{\scriptsize{1}}$
contained the ensembles for $\sigma_{\scriptsize\mbox{t}}^{\pm}$,
$\rho^{\pm}$ from \cite{PRD-110-030001-2024} is standard one. That
database is denoted as DB24 while the database taken into account
the above samples and results for
$\sigma_{\scriptsize\mbox{t}}^{+}$ at $\sqrt{s}=0.20$, 13 TeV from
STAR \cite{PLB-808-135663-2020} and ATLAS \cite{EPJC-83-441-2023}
respectively, for $\rho^{+}$ at $\sqrt{s}=13$ TeV from ATLAS
\cite{EPJC-83-441-2023} and TOTEM \cite{EPJC-79-785-2019} is
referred as DB24+. Influence of the TOTEM results for $\rho^{+}$
on the version of DB24+ was analyzed in details elsewhere
\cite{Okorokov-CPC-46-083105-2022}. Table \ref{tab:3} summarizes
the main features of the databases of experimental results
discussed in the present work for the set of the scattering
parameters $\mathcal{G}_{\scriptsize{1}}$.

The set of free parameters used within AQFT approach
$\mathcal{P}=\{\mathcal{P}^{i}\}_{i=1}^{9}=
\{\alpha_{+,-},\beta_{+,-},\gamma_{+,-},\delta_{+,-},\kappa\}$ can
be obtained from the fitting procedures of experimental data. One
can stress that each of $\sigma_{\scriptsize{\mbox{t}}}^{\pm}$ is
driven by the corresponding subset
$\mathcal{P}_{\pm}=\{\mathcal{P}^{i}_{\pm}\}_{i=1}^{4}=
\{\alpha_{\pm},\beta_{\pm},\delta_{\pm},\gamma_{\pm}\}$ of 4 free
parameters whereas $\rho^{\pm}$ depends on full set $\mathcal{P}$
of 9 free parameters due to interference term
$[\sigma_{\footnotesize\mbox{t}}
^{+}-\sigma_{\footnotesize\mbox{t}}^{-}]$ in (\ref{eq:2.5}). Thus,
experimental data for $\rho$-parameter for only one ($pp /
\bar{p}p$) scattering allow, in general, the derivation of values
for full set $\mathcal{P}$ of free parameters, i.e. for both $pp$
and $\bar{p}p$ reactions, within AQFT approach.

The energy dependence of the terms of
$\mathcal{G}_{\scriptsize{1}}$ formed from DB24 is approximated by
(\ref{eq:2.4}), (\ref{eq:2.5}) in the range $\sqrt{s} \geq
\sqrt{s_{\scriptsize{\mbox{min}}}}$, where
$\sqrt{s_{\scriptsize{\mbox{min}}}}$ is the low boundary of fit
range which is defined empirically. In the present work the wide
collection
$\bigl\{\sqrt{s_{\scriptsize{\mbox{min}},i}}\bigr\}_{i=1}^{16}=\{3,
5, 10, 15, 20, 25, 30, 40, 50, 60, 86.45, 100, 500, 10^{3}, 5
\times 10^{3}, 10^{4}\}$ GeV is used in order to detailed scan
$\sigma_{\scriptsize{\mbox{t}}}^{\pm}(s)$, $\rho^{\pm}(s)$ and
search for possible approach to the asymptotic
behavior\footnote{The set
$\bigl\{\sqrt{s_{\scriptsize{\mbox{min}},i}}\bigr\}_{i=1}^{16}$
considered here is the joined one for studies
\cite{Okorokov-CPC-46-083105-2022},
\cite{Okorokov-IJMPA-25-5333-2010,Okorokov-IJMPA-32-1750175-2017}
and the point $\sqrt{s_{\scriptsize{\mbox{min}},11}}=86.45$ GeV
corresponded the appearance of the first data point from
cosmic-ray measurements for $\sigma_{\scriptsize{\mbox{t}}}^{+}$
\cite{Okorokov-PAN-81-508-2018}.}. For the respective experimental
points, the total uncertainties are considered which are
calculated as
$\Delta^{2}_{\scriptsize{\mbox{tot}}}=\Delta^{2}_{\scriptsize{\mbox{stat}}}+
\Delta^{2}_{\scriptsize{\mbox{sys}}}$, where
$\Delta^{2}_{\scriptsize{\mbox{stat (sys)}}}$ stands for
statistical (systematic) errors.

\begin{table}[t!]
\centering \caption{Databases for the set of global scattering
parameters $\mathcal{G}_{\scriptsize{1}}$.}\label{tab:3}
\begin{tabular}{lcccc} \hline
\multicolumn{1}{c}{Database} & \multicolumn{4}{c}{Parameter from
the set $\mathcal{G}_{1}$} \\\cline{2-5}\rule{0pt}{12pt} &
$\sigma_{\scriptsize\mbox{t}}^{+}$&
$\sigma_{\scriptsize\mbox{t}}^{-}$ & $\rho^{+}$&
$\rho^{-}$ \rule{0pt}{12pt}\\
\hline
DB24        & ~~~\cite{PRD-110-030001-2024} & ~~~\cite{PRD-110-030001-2024}  & ~~~\cite{PRD-110-030001-2024} & ~~~\cite{PRD-110-030001-2024} \rule{0pt}{10pt}\\
DB24$_{1}$+ &
~~~\cite{PRD-110-030001-2024,PLB-808-135663-2020,EPJC-83-441-2023}
& ~~~--//-- & ~~~\cite{PRD-110-030001-2024,EPJC-83-441-2023} and
$\left.\langle
\rho^{pp}\rangle\right|_{\sqrt{s}=13\,\scriptsize{\mbox{TeV}}}$ \cite{Okorokov-CPC-46-083105-2022}& ~~~--//-- \rule{0pt}{10pt}\\
DB20$_{2}$+ & ~~~--//-- & ~~~--//-- & ~~~\cite{PRD-110-030001-2024,EPJC-79-785-2019,EPJC-83-441-2023} & ~~~--//-- \rule{0pt}{10pt}\\
\hline
\end{tabular}
\end{table}

It should be stressed that there are only few experimental points
for $\bar{p}p$ reaction at $\sqrt{s} > 1$ TeV. Moreover the data
for that interaction are absent at $\sqrt{s} > 5$ TeV at all.
Therefore the simultaneous fit for global scattering parameters is
made for $pp$ only at $\sqrt{s_{\scriptsize{\mbox{min}},15}}=5$
TeV and numerical values of terms of the full set $\mathcal{P}$
are derived due to (\ref{eq:2.5}) for $\rho^{+}(s)$. In DB24 there
are experimental results for
$\sigma_{\footnotesize\mbox{t}}^{+}(s)$ only at
$\sqrt{s_{\scriptsize{\mbox{min}},16}}=10$ TeV and, consequently,
simultaneous fit procedure reduces to the individual fit for $pp$
total cross section by (\ref{eq:2.4}). Similar study was made for
AQFT in our previous work \cite{Okorokov-PAN-81-508-2018}. Here
the individual fit for $\sigma_{\footnotesize\mbox{t}}^{+}(s)$ is
not considered. The simultaneous fit of $pp$ part of the set
$\mathcal{G}_{1}$ --
$\mathcal{G}_{+}=\{\mathcal{G}_{+}^{j}\}_{j=1}^{2}=
\{\sigma_{\scriptsize{\mbox{t}}}^{+},\rho^{+}\}$ -- can be made
for versions of DB24+ (Table \ref{tab:3}) at highest
$\sqrt{s_{\scriptsize{\mbox{min}},16}}=10$ TeV. This work is in
the progress. The features of DB24 and fit procedure at high
$\sqrt{s_{\scriptsize{\mbox{min}}}} \geq 1$ TeV, in particular
available $\bar{p}p$ data, should be taken into account for future
discussion.

\section{Fit results and discussion}\label{sec:4}

\begin{figure*}
\includegraphics[width=14.0cm,height=14.0cm]{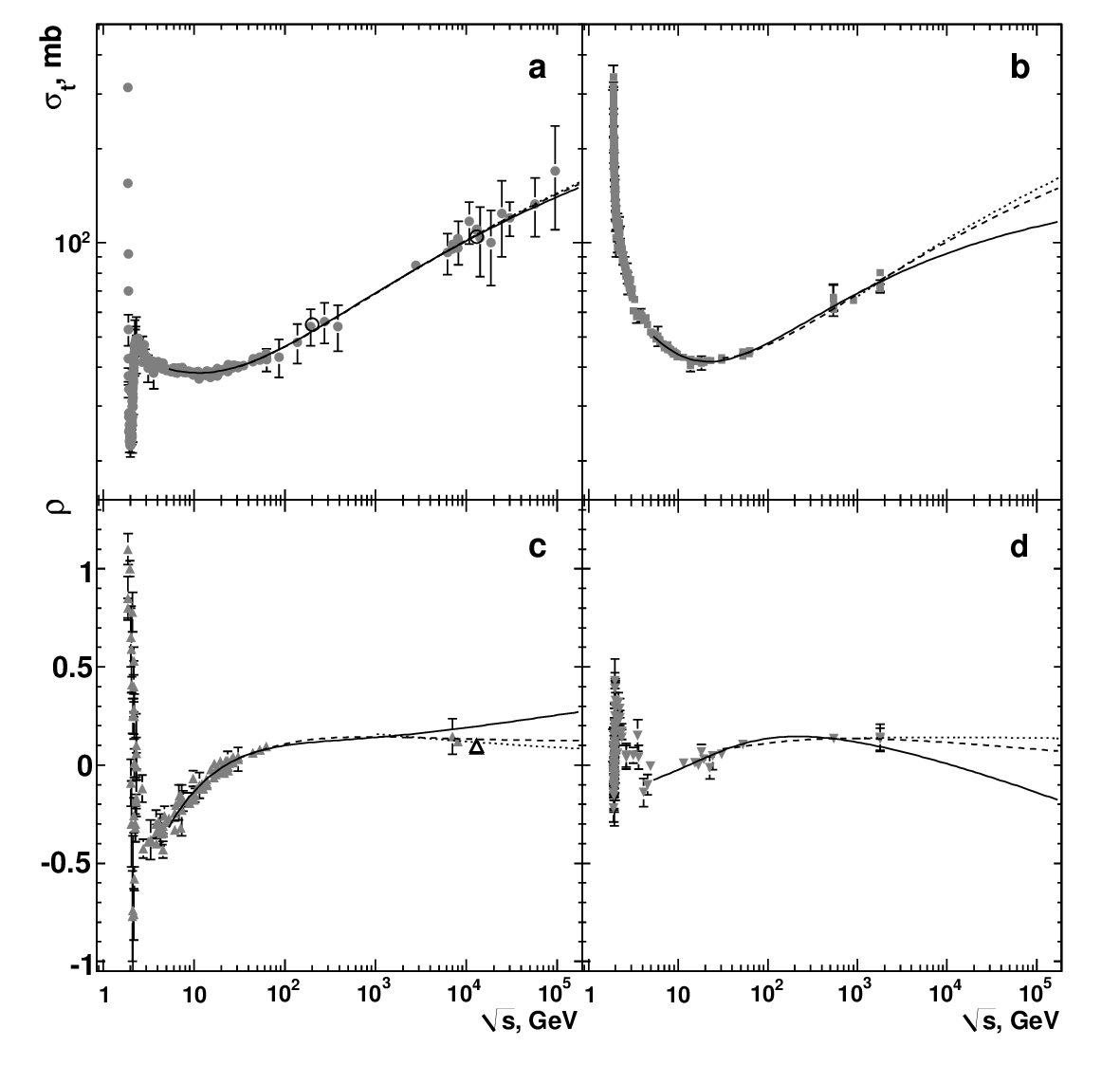}
\vspace*{8pt} \caption{The energy dependence for
$\sigma_{\scriptsize{\mbox{t}}}^{+}$ (a),
$\sigma_{\scriptsize{\mbox{t}}}^{-}$ (b), $\rho^{+}$ (c),
$\rho^{-}$ (d) and results of simultaneous fits of all four
parameters by equations (\ref{eq:2.4}), (\ref{eq:2.5}). Solid
points are from DB24 and these points are included in the fitted
sample, open points are additional for DB24 and they are not
included in the fitted data sample (detailed explanation -- see in
text). The solid line corresponds to the fit at
$\sqrt{s_{\scriptsize{\mbox{min}}}}=5$ GeV, the dashed line -- at
$\sqrt{s_{\scriptsize{\mbox{min}}}}=30$ GeV and dotted line -- at
$\sqrt{s_{\scriptsize{\mbox{min}}}}=1$ TeV.}\label{fig:1}
\end{figure*}

In the present work simultaneous fits are made by equations
(\ref{eq:2.4}), (\ref{eq:2.5}) for the energy dependence of global
scattering parameters from the set $\mathcal{G}_{1}$ for DB24. The
numerical results of the simultaneous fits are shown in Table
\ref{tab:4} for low boundaries $s_{\scriptsize{\mbox{min}},i}$,
$i=1-15$ of approximated energy range\footnote{The first value in
the cell of Table \ref{tab:4} corresponds to the model parameter
with plus sign, i.e. for $pp$, and second value -- to the model
parameter with minus sign, i.e. for $\bar{p}p$; $\chi^{2}_{1}
\equiv \chi^{2}$/n.d.f.}. Experimental data from DB24 together
with fit curves are shown in Fig. \ref{fig:1}. Also additional
points from DB24$_{2}$+ points are shown for completeness, but as
indicated above the points from DB24 are only fitted in Fig.
\ref{fig:1}. Here solid line correspond to the fits at
$\sqrt{s_{\scriptsize{\mbox{min}},2}}=5$ GeV, dashed curves are
fits at $\sqrt{s_{\scriptsize{\mbox{min}},7}}=30$ GeV and dotted
lines are for $\sqrt{s_{\scriptsize{\mbox{min}},14}}=1$ TeV.

\begin{table}[t!]
\centering \caption{Results for simultaneous fitting by equations
(\ref{eq:2.4}), (\ref{eq:2.5}) of $\mathcal{G}_{1}(s)$ for DB24.}
\label{tab:4}
\begin{tabular}{lcccccc}\hline
\multicolumn{1}{l}{$i$} &
\multicolumn{6}{c}{Parameter} \\
\cline{2-7} \rule{0pt}{12pt} & $\alpha_{\pm} \times 10^{3}$ &
$\beta_{\pm} \times 10^{3}$ & $-\gamma_{\pm} \times 10^{3}$ &
$\delta_{\pm}$, mb & $\kappa$, mb & $\chi^{2}_{1}$
\\\hline 
1 & $20.0 \pm 1.3$& $12.84 \pm 0.29$ & $5886.1 \pm 2.0$ & $51.51 \pm 0.22$ & $42 \pm 5$   & 1.88 \\
  & $76.4 \pm 2.1$& $32.0 \pm 0.8$   & $5682 \pm 4$   & $124.9 \pm 1.4$  &              & \\
2 & $18.1 \pm 2.3$& $12.4 \pm 0.6$   & $5890 \pm 5$   & $51.0 \pm 0.7$   & $-46 \pm 14$ & 1.13 \\
  & $52 \pm 4$    & $22.4 \pm 1.2$   & $5763 \pm 8$   & $96.0 \pm 2.5$   &              & \\
3 & $16 \pm 4$   & $11.9 \pm 1.1$   & $5895 \pm 11$   & $50.3 \pm 1.9$   & $-110 \pm 30$ & 1.03 \\
  & $36 \pm 7$    & $17.3 \pm 2.0$   & $5815 \pm 18$   & $79 \pm 5$       &              & \\
4 & $14 \pm 6$   & $11.4 \pm 1.3$   & $5900 \pm 15$   & $49.6 \pm 2.7$   & $-42 \pm 29$ & 1.00 \\
  & $20 \pm 9$    & $12.8 \pm 2.1$   & $5864 \pm 20$   & $65 \pm 5$       &              & \\
5 & $22 \pm 6$   & $13.6 \pm 1.7$   & $5873 \pm 21$   & $55 \pm 5$       & $35 \pm 13$  & 0.92 \\
  & $25 \pm 9$    & $14.3 \pm 2.2$   & $5842 \pm 25$   & $72 \pm 7$       &              & \\
6 & $15 \pm 5$   & $11.8 \pm 1.3$   & $5895 \pm 16$   & $50 \pm 3$       & $450 \pm 280$ & 0.84 \\
  & $19 \pm 10$   & $12.8 \pm 2.1$   & $5853 \pm 23$   & $71 \pm 7$       &              & \\
7 & $13 \pm 8$   & $11.0 \pm 1.9$   & $5908 \pm 24$   & $47 \pm 5$       & $420 \pm 290$ & 0.86 \\
  & $20 \pm 13$    & $13.0 \pm 2.9$   & $5850 \pm 40$   & $72 \pm 13$      &              & \\
8 & $12.5 \pm 1.6$ & $11.07 \pm 0.29$ & $5901.6 \pm 2.1$ & $49.6 \pm 1.4$ & $1000 \pm 120$ & 1.02 \\
  & $9 \pm 5$     & $10.2 \pm 0.6$   & $5899 \pm 5$   & $56 \pm 3$       &              & \\
9 & $8 \pm 3$    & $9.8 \pm 0.5$    & $5931 \pm 4$   & $41.9 \pm 2.2$   & $140 \pm 30$ & 1.08 \\
  & $1.3 \pm 0.8$ & $8.64 \pm 0.06$  & $5925.8 \pm 1.0$ & $48.6 \pm 0.4$   &              & \\
10 & $1.5 \pm 0.8$& $8.41 \pm 0.16$& $5946 \pm 8$   & $40.3 \pm 2.6$   & $2000 \pm 600$ & 0.97 \\
  & 0.0 (fixed)   & $8.3 \pm 0.5$    & $5933 \pm 19$   & $47 \pm 6$       &              & \\
11 & $1.94 \pm 0.15$ & $8.51 \pm 0.14$ & $5946.6 \pm 1.9$ & $39 \pm 4$  & $2000 \pm 700$ & 0.99 \\
  & 0.0 (fixed)   & $8.3 \pm 0.4$    & $5926.4 \pm 2.8$ & $50 \pm 4$       &              & \\
12 & $1.95 \pm 0.14$ & $8.50 \pm 0.16$ & $5946.5 \pm 1.8$ & $39 \pm 3$  & $2100 \pm 700$ & 1.02 \\
  & 0.0 (fixed)   & $8.4 \pm 0.4$    & $5926.4 \pm 2.7$ & $50 \pm 4$       &              & \\
13 & $2.4 \pm 0.5$ & $8.60 \pm 0.12$ & $5944.4 \pm 2.1$ & $40 \pm 3$      & $2300 \pm 700$ & 1.17 \\
  & 0.0 (fixed)   & $8.3 \pm 0.4$    & $5934 \pm 4  $   & $46 \pm 4$      &              & \\
14 & 0.0 (fixed) & $7.1 \pm 0.7$     & 6000 (fixed)     & $26 \pm 7$      & $1800 \pm 800$ & 1.25 \\
  & --//--        & $7.6 \pm 1.2$    & --//--           & $22 \pm 8$       &              & \\
15 & 0.0 (fixed) & $9.69 \pm 0.06$   & 6000 (fixed) & 0.0 (fixed) & 0.0 (fixed)    & 0.99 \\
  & --//--        & $10.3 \pm 0.4$   & --//--              & --//--           &              & \\
\hline
\end{tabular}
\end{table}

As it is seen suggested approximation functions (\ref{eq:2.4}),
(\ref{eq:2.5}) provide reasonable fit qualities at smallest
$\sqrt{s_{\scriptsize{\mbox{min}},1}}=3$ GeV and, may be, at
$\sqrt{s_{\scriptsize{\mbox{min}},14}}=1$ TeV; statistically
acceptable $\chi^{2}/\mbox{n.d.f.}$ are obtained at any other
$\sqrt{s_{\scriptsize{\mbox{min}}}}$. Thus, as previously
\cite{Okorokov-IJMPA-25-5333-2010,Okorokov-IJMPA-32-1750175-2017}
the parameterizations for the terms of the set $\mathcal{G}_{1}$
deduced within AQFT show good agreement with available
experimental data. The recent RHIC \cite{PLB-808-135663-2020} and
LHC \cite{EPJC-83-441-2023,EPJC-79-785-2019} results shown by open
symbols in Fig. \ref{fig:1} agree well with general trends for
global $pp$ scattering parameters (Fig. \ref{fig:1}a, c)
especially for $\sigma_{\scriptsize{\mbox{t}}}^{+}$ (Fig.
\ref{fig:1}a). Therefore, one can expect the new data with respect
to the DB24 will negligibly effect on the values of fit
parameters, especially for high $s_{\scriptsize{\mbox{min}}}$.
This qualitative hypothesis is quite confirmed by many results of
our previous studies
\cite{Okorokov-CPC-46-083105-2022,Okorokov-IJMPA-25-5333-2010,
Okorokov-IJMPA-32-1750175-2017,Okorokov-PAN-81-508-2018} made for
earlier different versions of experimental database for terms of
the set $\mathcal{G}_{1}$.

The values of all terms of the set $\mathcal{P}$ from Table
\ref{tab:4} obtained for DB24 at low and intermediate
$\bigl\{\sqrt{s_{\scriptsize{\mbox{min}},i}}\bigr\}_{i=1}^{7}$
mostly coincide with the values of corresponding fit parameters
from \cite{Okorokov-IJMPA-25-5333-2010,
Okorokov-IJMPA-32-1750175-2017} within errors for any
$s_{\scriptsize{\mbox{min}}}$ indicated above. Fit qualities are
(very) close to the published ones
\cite{Okorokov-IJMPA-25-5333-2010, Okorokov-IJMPA-32-1750175-2017}
at equal
$\bigl\{\sqrt{s_{\scriptsize{\mbox{min}},i}}\bigr\}_{i=1}^{7}$. As
consequence all features of the fit curves in Fig. \ref{fig:1}
shown by solid ($\sqrt{s_{\scriptsize{\mbox{min}},2}}=5$ GeV) and
dashed ($\sqrt{s_{\scriptsize{\mbox{min}},7}}=30$ GeV) lines agree
with detailed discussion for our previous works
\cite{Okorokov-IJMPA-25-5333-2010,Okorokov-IJMPA-32-1750175-2017}.
Detailed analysis shows that values of some free parameters
coincide with null or / and its asymptotic levels for modified
Froissart--Martin limit within uncertainties starting from
$\sqrt{s_{\scriptsize{\mbox{min}},10}}=60$ GeV already. In these
cases corresponding experimental samples from DB24 are fitted at
fixed values of such parameters (Table \ref{tab:2}). Thus the
functional behavior of $\sigma_{\scriptsize{\mbox{t}}}^{\pm}(s)$
approaches consecutively and gradually to the modified
Froissart--Martin limit with growth of
$\sqrt{s_{\scriptsize{\mbox{min}}}}$ starting with high value
$\sqrt{s_{\scriptsize{\mbox{min}},14}}=1$ TeV. As it is seen from
Table \ref{tab:4} the following relations are valid
$\sigma_{\scriptsize{\mbox{t}}}^{\pm}(s)=\beta_{\pm}\sigma_{\scriptsize{\mbox{FM}}}$
and $\rho^{\pm}(s)=\rho_{\scriptsize{\mbox{FM}}}$ at highest
boundary for simultaneous fit
$\sqrt{s_{\scriptsize{\mbox{min}},15}}=5$ TeV. Moreover
$\sigma_{\scriptsize{\mbox{t}}}^{+}(s)=\sigma_{\scriptsize{\mbox{t}}}^{-}(s)$
due to coincidence of $\beta_{+}$ and $\beta_{-}$ within better
than 1.33 standard deviation (Table \ref{tab:4}). It means that
any formulations of Pomeranchuk theorem are valid and all global
scattering parameters considered here reach the modified
Froissart--Martin limit at least in functional sense because of
$\beta_{\pm} \ll \beta_{\scriptsize{\mbox{FM}}}$. The result
$\sqrt{s_{a,\scriptsize{\mbox{FM}}}}=5$ TeV can be considered as
qualitative indication only for the onset of asymptotic region for
global scattering parameters with point of view of the theorems
under consideration because of the reasons discussed above,
namely, (i) limited data sample for $pp$ collisions, especially,
for $\rho^{+}(s)$, (ii) availability of experimental results for
$\rho^{+}(s)$ in (very) narrow energy range $\sqrt{s}=7-8$ TeV,
and (iii) absence of $\bar{p}p$ data at all. All of these can
influence on the robustness and stability for values of terms of
the set $\mathcal{P}$ and, consequently, on physical conclusions.
On the other hand, the result of the simultaneous fit at
$\sqrt{s_{\scriptsize{\mbox{min}},15}}=5$ TeV agree qualitatively
with the previous individual analysis of
$\sigma_{\footnotesize\mbox{t}}^{+}(s)$ at
$\sqrt{s_{\scriptsize{\mbox{min}},16}}=10$ TeV
\cite{Okorokov-PAN-81-508-2018} provided the functional behavior
of this quality close to the (modified) Froissart--Martin limit.

The large quantitative discrepancy between
$\sigma_{\scriptsize{\mbox{t}}}^{\pm}$ and
$\sigma_{\scriptsize{\mbox{FM}}}$ observed here and in previous
studies
\cite{Okorokov-IJMPA-25-5333-2010,Okorokov-IJMPA-32-1750175-2017,Okorokov-PAN-81-508-2018}
even for multi-TeV region may indicate on the necessity of some
novel dynamical mechnisms for $pp / \bar{p}p$ scattering which
could be provide significant increase of total cross sections,
i.e. $\beta_{\pm}$ values within AQFT approach. In general,
Bose--Einstein condensation (BEC) can provide the noticeable
growth of multiplicity of secondary pions in nucleus--nucleus
collisions at sufficiently high energies
\cite{Okorokov-PAN-87-172-2024}. Therefore, in general, BEC can
affect the behavior of cross sections. Although this effect was
not obtained for $pp$ up to $\sqrt{s} \lesssim 1$ PeV it can
appears at larger energies. Then the BEC could be provide, for
instance, the sharp increase of total cross section at certain
$\sqrt{s}$ or in (very) narrow range of collision energy like to
behavior of multiplicity of secondary pions
\cite{Okorokov-PAN-87-172-2024} without violation of
Froissart--Martin limit on future functional behavior of
$\sigma_{\scriptsize{\mbox{t}}}(s)$. The scenario is similar to
that for some parameter at 1st order phase transition with
discontinuity and jump to the new (significantly higher) level of
values. Thus, perhaps, BEC can be suggested as one of the possible
dynamical mechanisms which may provide the approach to (modified)
Froissart--Martin limit quantitatively in "truly asymptotic
region" which onset can be expected, in general, up to grand
unified theory (GUT) energy domain in order of magnitude, i.e.
$\sqrt{s_{a}} \gtrsim 10^{12}-10^{13}$ GeV
\cite{Okorokov-IJMPA-33-1850077-2018}. This qualitative hypothesis
request the future development and detailed study.

\begin{figure*}
\includegraphics[width=14.0cm,height=14.0cm]{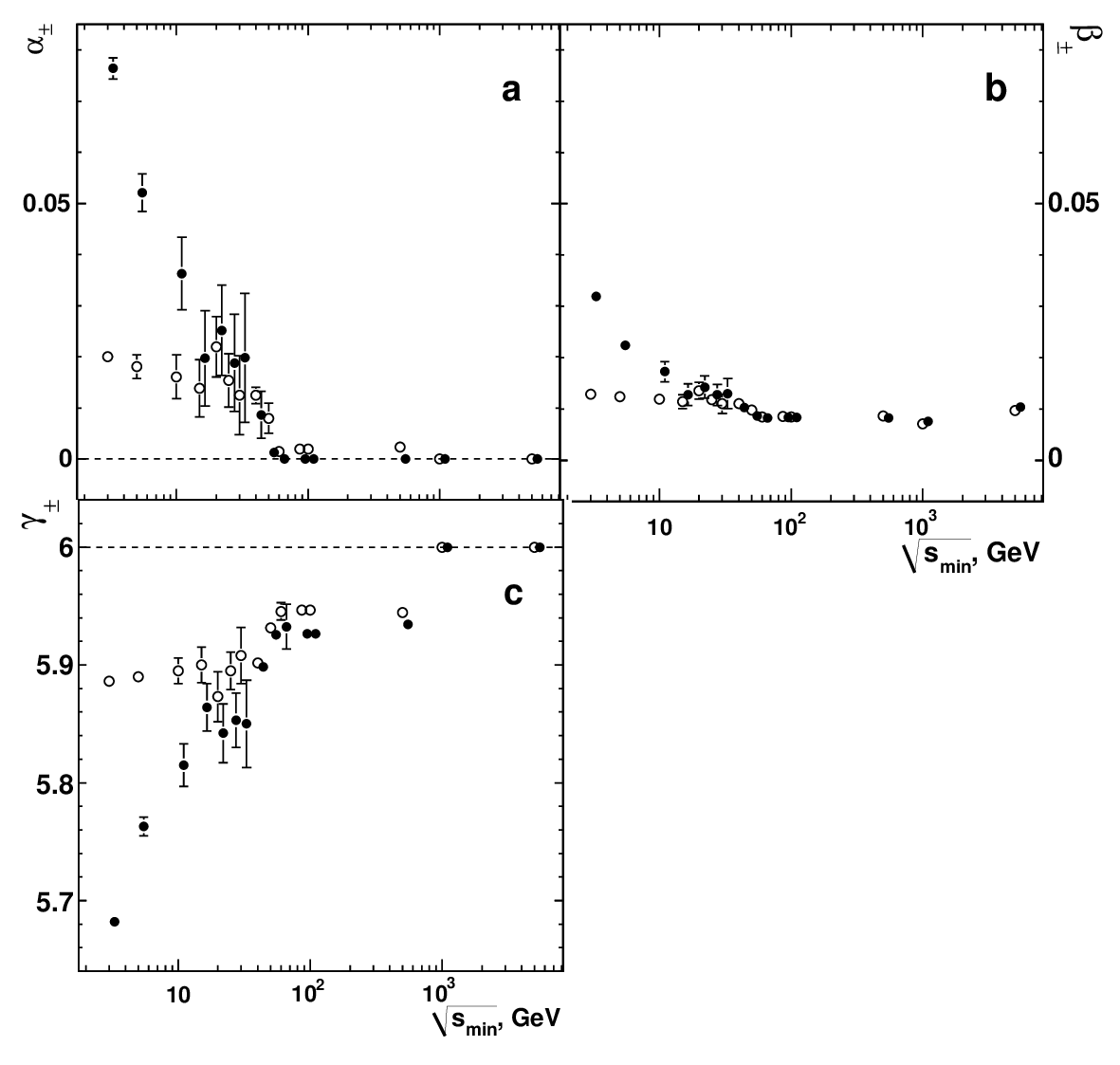}
\vspace*{8pt} \caption{Dependence on
$\sqrt{s_{\scriptsize{\mbox{min}}}}$ for fit parameters
$\alpha_{\pm}$ (a), $\beta_{\pm}$ (b) and $\gamma_{\pm}$ (c). Open
points are for $pp$, solid ones -- for $\bar{p}p$. Dashed lines on
(a, c) correspond to the asymptotic levels for $\alpha_{\pm}$ and
$\gamma_{\pm}$. The points for $\bar{p}p$ are shifted on right for
more comfortable and clear view.}\label{fig:2}
\end{figure*}

Fig. \ref{fig:2} shows the dependence of $\alpha_{\pm}$ (a),
$\beta_{\pm}$ (b) and $\gamma_{\pm}$ (c) on the low boundary of
fit range $\sqrt{s_{\scriptsize{\mbox{min}}}}$. Values of
parameters $\alpha_{\pm}$ show sharp decrease at
$\sqrt{s_{\scriptsize{\mbox{min}}}} \geq 30$ GeV approaching to
the asymptotic level at $\sqrt{s_{\scriptsize{\mbox{min}}}} \geq
60$ GeV already (Fig. \ref{fig:2}a) whereas the similar statement
is valid for $\gamma_{\pm}$ at $\sqrt{s_{\scriptsize{\mbox{min}}}}
\geq 1$ TeV (Fig. \ref{fig:2}c). The parameter $\beta_{\pm}$ shows
weaker dependence on $\sqrt{s_{\scriptsize{\mbox{min}}}}$
especially in the range $\sqrt{s_{\scriptsize{\mbox{min}}}}
\gtrsim 50-60$ GeV (Fig. \ref{fig:2}b). Some increase is observed
for $\beta_{\pm}$ at these energies but $\beta_{\pm} /
\beta_{\scriptsize{\mbox{FM}}} \lesssim 10^{-2}$ even at highest
$\sqrt{s_{\scriptsize{\mbox{min}},15}}$ under consideration. The
relations between free parameters of AQFT model agree with
requests of Table \ref{tab:1} starting at
$\sqrt{s_{\scriptsize{\mbox{min}},14}}=1$ TeV, i.e. AQFT allows
the validity of both formulations of Pomeranchuk theorem at
energies $\sqrt{s} \geq \sqrt{s_{a,\scriptsize{\mbox{P}}}}=1$ TeV.

The results obtained by simultaneous fitting at
$\sqrt{s_{\scriptsize{\mbox{min}}}} \geq 1$ TeV are correct from
the point of view of the methods of physical data analysis.
However, it seems that aforementioned (very) limited sample of
$\bar{p}p$ experimental data at $\sqrt{s} > 1$ TeV and the
complete absence of such data at $\sqrt{s} > 2$ TeV allow only
qualitative and preliminary conclusions about the presence of
indications of the fulfillment of Pomeranchuk's theorem at
energies above 1 TeV and the achievement of the Froissart--Martin
limiting behavior by the $\sigma_{\scriptsize{\mbox{t}}}^{\pm}$ in
the functional sense at $\sqrt{s} \gtrsim 5$ TeV. Experimental
data on $\bar{p}p$ scattering are very important in the multi-TeV
region for more precise and definitive physical conclusions.
However, at present, as it is known, projects to create a collider
with $\bar{p}$ beam for energies comparable to the nominal energy
of the LHC design, i.e. at $\sqrt{s} \sim 10$ TeV, are not being
considered even in the medium-term perceptive in time. It seems
that the development and implementation of a project for such an
$\bar{p}p$ collider would be important from the point of view of
the physics of strong interactions in general, and, in particular,
for the study of global scattering parameters in $\bar{p}p$
interactions in a new energy region.

\section{Summary}\label{sec:5}

The analytic parameterizations suggested within AQFT provide the
quantitative description of energy dependence of global scattering
parameters for $pp$ and $\bar{p}p$ scattering with robust values
of fit parameters for recent DB24. The wide set of low boundary
for fit domain is used. It allows the detailed scan of behavior of
total cross sections and $\rho$-parameter in dependence of
collision energy. Value of free parameters for AQFT model agree
well with our previous works at low and intermediate
$s_{\scriptsize{\mbox{min}}}$. Based on the present scan on
$s_{\scriptsize{\mbox{min}}}$ extended to the high-energy region
the numerical estimations are derived for the onsets of energy
regions in which Pomeranchuk theorem and / or Froissart--Martin
one is valid. Phenomenological parameterizations within AQFT and
available experimental data allow the validity of both
formulations of Pomeranchuk theorem at energies $\sqrt{s} \geq 1$
TeV. It is observed that all global scattering parameters
considered here reach the modified Froissart--Martin limit at
least in functional sense at $\sqrt{s} \geq 5$ TeV. But the
numerical values of total cross sections are much smaller of
corresponding asymptotic levels. Perhaps, the investigation of
bosonic condensation will shed new light on the nature of this
quantitative discrepancy. It seems that the development and
building of new $\bar{p}p$ collider for study of multi-TeV energy
region would be very useful, in particular, for physics of strong
interaction processes at large distance and for direct search for
the onset of asymptotic region for global scattering parameters.
\newpage
\newpage

\end{document}